\documentclass[12pt]{article}
\usepackage{amsmath,amsthm,amsfonts,amssymb,amscd}
\usepackage{graphics}
\usepackage[latin1]{inputenc}

\begin{document}

\newcommand{\la}{{\lambda}}
\newcommand{\ro}{{\rho}}
\newcommand{\po}{{\partial}}
\newcommand{\ov}{\overline}
\newcommand{\re}{{\mathbb{R}}}
\newcommand{\nb}{{\mathbb{N}}}
\newcommand{\Z}{{\mathbb{Z}}}
\newcommand{\Uc}{{\mathcal U}}
\newcommand{\gc}{{\mathcal G}}
\newcommand{\hc}{{\mathcal M}}
\newcommand{\fc}{{\mathcal F}}
\newcommand{\dc}{{\mathcal D}}
\newcommand{\al}{{\alpha}}
\newcommand{\vr}{{\varphi}}
\newcommand{\om}{{\omega}}
\newcommand{\La}{{\Lambda}}
\newcommand{\be}{{\beta}}
\newcommand{\te}{{\theta}}
\newcommand{\Om}{{\Omega}}
\newcommand{\ve}{{\varepsilon}}
\newcommand{\ga}{{\gamma}}
\newcommand{\Ga}{{\Gamma}}
\newcommand{\zb}{{\mathbb{Z}}}
\def\sen{\operatorname{sen}}
\def\Ker{\operatorname{{Ker}}}
\newcommand{\bc}{{\mathcal B}}
\newcommand{\lc}{{\mathcal L}}

\def\Lg{\bigg\langle}
\def\Rg{\bigg\rangle}
\renewcommand{\lg}{{\langle}}
\newcommand{\rg}{{\rangle}}
\newcommand{\lV}{{\left\Vert \right.}}
\newcommand{\rV}{{\left.\right\Vert}}
\def\H{\mathcal{H}}
\def\bn{\mathbb{N}}
\def\bn{\mathbb{N}}

\def\ro{\rho}
\def\bq{\mathbb{Q}}
\def\bz{\mathbb Z}
\def\supp{\operatorname{Supp}}
\def\amp{\operatorname{amp}}
\def\tor{\operatorname{Tor}}
\def\ker{\operatorname{Ker}}
\def\hom{\operatorname{Hom}}
\def\ext{\operatorname{Ext}}
\def\Tr{\operatorname{Tr}}
\def\det{\operatorname{det}}
\def\inn{\operatorname{in}}
\def\out{\operatorname{out}}
\def\cov{\operatorname{cov}}
\def\cosh{\operatorname{cosh}}
\def\tanh{\operatorname{tanh}}
\def\bind{\operatorname{bind}}

\newcommand{\cb}{{\mathbb C}}

\def\esimo{${}^{\text{\b o}}$}


\centerline{\huge\bf Some Comments on Infinities on}

\vskip .2in

\centerline{\huge\bf    Quantum Field Theory:}

\vskip .2in

\centerline{\huge\bf A Functional Integral Approach}

\thispagestyle{empty}

\vskip .5in

\centerline{{\sc Luiz C.L. Botelho}\footnote{Presently visiting the Math. Department University of Arizona under a CNPq Felowship.}}

\vskip .2in

\centerline{Departamento de Matemática Aplicada,}

\centerline{Instituto de Matemática, Universidade Federal Fluminense,}

\centerline{Rua Mario Santos Braga}

\centerline{24220-140, Niterói, Rio de Janeiro, Brazil}

\centerline{Correspondence should be addressed to }

\centerline{e-mail: botelho.luiz@ig.com.br}

\vskip .5in

\begin{abstract}
We analyze on the formalism of probabilities measures-functional integrals on function space, the problem of infinities on Euclidean field theories. We also clarify and generalize ours previous published studies on the subject.
\end{abstract}

\section{Introduction}

The subject of infinities of Euclidean Quantum Field Theories, or in general Minkowskian Quantum Field Theories, has been well grounded with practical Calculations on Quantum Physics since its full inception on 1950 years ([1]).

The purpose of this research note is to study the nature of ultraviolet infinities on Euclidean Quantum Field Path integral through an analitically regularized, mathematical rigorously path integrals. This study is presented in Section 1. On the Section 2, we clarify our previous studies on the subject ([2], [7]) by analyzing in detail all those estimates leading to a correct understanding of the well-known problem of coupling constant renormalization on QFT´s in a two-dimensional scalar interacting field model. In Section 3, mainly of pedagogical purpose, we present a mathematically detailed construction of the Wiener Kac functional measure. In Section 4, we analyze the mathematical structure of the so called Feynman geometrodynamical propagation on Euclidean QFT, by using the mathematical rigorous results of Section 3. 

In several appendixes, we add several calculations useful to understand the bulk of our research note.

\section{Infinities on Quantum Field Theory on the Functional Integral Formalism}

Euclidean quantum fields are expected to be mathematically defined objects by the analytic continuation of Minkowskian self-adjoints quantum fields to imaginary time, where this precise analytic continuation is well expressed through the famous Bargmann-Hall-Wightman theorem ([1]). It is thus expected that Euclidean self-adjoints quantum field operators on $R^D$ should be naturally constructed (from a rigorous mathematical point of view) from a probability measure $d\mu(\phi)$ on the L. Schwartz Tempered Districutions $S'(R^D)$. It is also expected that the following functional, the called theory's generating functional $Z(j)\in C(S(R^D),R)$ should furnishes the bridge of such time analytic extension though the identity
\begin{equation}
\langle \Om_{\text{vac}}^{\text{eucl}} |\exp i\phi(j)| \Om_{\text{vac}}^{\text{eucl}} \rangle = Z(j). \tag{1}
\end{equation}
Here $\phi(j)=\langle\phi,j\rangle$ is the usual canonical pairing between the distribution valued euclidean field operator $\phi$ (take  from here on as a neutral scalar field for simplicity of our exposition).

Since eq(1) is a Bochner-Martin positive definite functional on $C(S(R^D),R)$ one can apply the Minlos Theorem to represent eq(1) by means of a probability measure on $S'(R^D)$
\begin{equation}
Z(j)=\int_{S'(R^D)} d\mu(\phi) \exp i \phi(j).\tag{2}
\end{equation}

For example, massive free scalar fields on $R^D$ given by the explicity generating functional below (exponential of a continuous bilinear form on $S(R^D)$)
\begin{equation}
Z(j)=\exp \left( - \frac12 T_{(-\Delta+m^2)^{-1}j}(j) \right). \tag{3}
\end{equation}
Here our $T$ (continuous application) is explicited by the formula
\begin{align*}
T\colon (S(R^D), \text{ strong topology}) & \to (S'(R^D), \text{ weak topology}) \\
j & \to T_{(-\Delta+m^2)^{-1} j} \tag{3-a}
\end{align*}
where for $f\in S(R^D)$, we have
\begin{equation}
T_{(-\Delta+m^2)^{-1}j} (f) = \int_{R^D}((-\Delta+m^2)^{-1}j)(x) f(x)d^D x . \tag{3-b}
\end{equation}
As a consequence of above exposed, one simple application of Minlos-Bochner Theorem to eqs(3-a)/(3-b) give us the massive free scalar euclidean generating functional
\begin{align*}
Z(j) &= \exp\left( - \frac12 T_{(-\Delta+m^2)^{-1}j}(j) \right) \\
&= \exp\left( -\frac12 \langle j,(-\Delta+m^2)^{-1}j \rangle_{L^2(R^D)} \right) \\
&= \int d_{(-\Delta+m^2)^{-1}} \mu(\phi)(\exp i \phi(j)). \tag{3-c}
\end{align*}

Interacting quntum field theories as perturbation around massive scalar fields, by theirs turn are also defined by an absolutely continuous measure $d\nu(\phi)$ in relation to previously euclidean massive free scalar field
$d_{(-\Delta+m^2)^{-1}}\mu(\phi)$, which in a rigorous way would be writen as of as
\begin{equation}
d_{(\ve)}\nu(\phi)=(d_{(-\Delta+m^2)^{-1}} \mu(\phi)) (e^{-g V(\phi * \rho_\ve)}) \tag{4}
\end{equation}
where $(\phi * \rho_\ve)$ denotes the field sampling in eq(3-c) in its regularized form (which are $C^\infty(R^N)$ functions).

Classifically $gV(\phi)$ is a non-linear function of the $C^\infty$-regularized distribution $(\phi * \rho_\ve)$ and $g$ is the bare theory's coupling constant.

Note that trying to formalize rigorously eq(4) for $\ve\to0$ is meaningless, since one expect that if these probabilistic distributional field configurations were described by locally integrable functions, they would be infinite almost everywhere on $R^D$. It is no difficult to expect such behaviour for scalar neutral free euclidean fields since (for $D>1$)\footnote{See Appendix B}

\begin{align*}
\int d_{(-\Delta+m^2)^{-1}} \mu(\phi) \phi^2(x) &= \lim_{x\to y} \int d_{(-\Delta+m^2)^{-1}} \mu(\phi)(\phi(x)\phi(y))\\
&= \frac1{(2\pi)^{D/2}} \int d^D k \frac1{k^2+m^2}=+\infty \tag{5}
\end{align*}

\medskip

As a consequence it does not make sense to consider interactiong euclidean QFT interactions directly from their classical counterpart Lagrangians as usually done in Physics textbooks. For instance, the functional below it holds true the triviality relation on the mentioned field measure space.
\begin{equation}
\exp \left\{ -\frac\la{4!} \int \phi^4(x) d^D x \right\} =0 \quad \text{ a.e. on }\quad (S'(R^D),d_{(-\Delta+m^2)^{-1}} \mu(\phi)). \tag{6}
\end{equation}

For interacting field theories as defined as perturbation such ``triviality''\, behavior is also expected.

In order to overcome such mathematical problems, it is used ``regularized'' forms of the QFT, which in many cases, although useful for computations, they shadow on the famous problems of infinities or the usual Feynman-Dhyson perturbative scheme (intrinsic for defining interaction field path measures around free field path integrals).

Let us now propose a new form of regularizing euclidean quantum field theories by considering suitable regularization on the Kinetic free operator eq(3-b).

We have the following basic theorem of ours ([2]).

\vskip .2in

\noindent
{\bf Theorem 1.} {\it Let us consider the $\al$-power of the Laplacean operator acting on $H^{2\al}(R^D)$. Let us consider also a compact domain $\Om\subset R^D$ and $\chi_\Om(\tau)$ its characteristic function 
$$
\left(\chi_\Om(x)= \begin{cases}
1 \quad x\in \Om \\
0 \quad x\notin \Om
\end{cases} \right)
$$
then one has:

\noindent
{\rm 1 --} the operator $\lc_{(\al,\Om,m)}$ here defined by its inverse which is given by an integral operator with kernel
\begin{equation}
\lc_{\al,\Om,m}^{-1}(x,y)=\chi_\Om(x) [(-\Delta)^\al+m^2]^{-1}(x,y) \chi_\Om(y) \tag{7}
\end{equation}
is such that it defines a probability measure on the functional space $L^2(R^D)$ for the parameter range $\al> D/2$
\begin{align*}
Z(j) &= \exp \left\{ -\frac12 \langle j, \lc_{(\al,\Om,m)}^{-1} j\rangle_{L^2(R^D)} \right\} \\
&= \int_{L^2(R^D)}  \left( d_{\lc^{-1}(\al,\Om,m)} \mu(\phi) \exp i \langle \phi,j \rangle \right).
\tag{8}
\end{align*}
}

\vskip .2in

\noindent
{\bf Proof:}
We note that $\lc_{\al,\Om,m}^{-1}$ is a positive definite trace class operator on $L^2(R^D)$ for $\al>\frac D2$ (ref. ([2])). Since
\begin{equation}
Tr_{L^2(R^D)} \{ \lc_{(\al,\Om,m)}^{-1} \} = \frac{\text{vol}(\Om)}{(2\pi)^D} \left[ \int_{R^D} \frac{d^Dk}{k^{2\al}+m^2} \right] < \infty \quad \text{ if }\quad \al>\frac D2. \tag{9}
\end{equation}

Now the result on the support of the probability measure given by eq(8) as given by the $L^2(R^D)$ (or $L^2(\Om)$) space is a direct consequence of the Minlos-Bochner theorem on integration theory on Hilbert Spaces setting ([2] -- appendix). \qed

\vskip .2in

The point now in consider finite volume analitically regularized free scalar QFT's is that one can handle directly non-trivial interactions (super renormalizable QFT's)
since the field condigurations are now the usual real measurable point $L^2(\Om)$ functions instead of L. Schwartz distributions.

For instance, we have the following theorem:

\vskip .2in

\noindent
{\bf Theorem 2.} {\it  The exponential cut-off euclidean $P_2(\vr)$ interaction (defined explicitly below) is well defined on the $(d_{\lc_{(\al,\Om,m)}^{-1}} \mu(\phi), L^2(\Om))$ probability functional space for $\al>\frac D2$
\begin{equation}
V_\Om(\vr)=\int_{\Om\subset R^D} \exp(-\delta\vr^2(x)) \left[ \sum_{\substack{j=0 \\ j=\text{even}}}^{2k} \frac{\la_j}{j'} \vr^j(x) \right] d^Dx \tag{10}
\end{equation}
here $\delta>0$, $\la_j\ge 0$, for $j=0,2,4,\dots,2k$.
}

\vskip .2in

\noindent
{\bf Proof:}
An immediate consequence of the fact that if $\vr(x)=+\infty$ for $\vr\in L^2(\Om)$ and for some point $x\in\Om$, then $V(\vr(x))=0$, which means that $e^{-V(\phi)}=1$. As a result one has the upper bound for all 
\begin{equation}
\phi\in L^2(\Om)=\supp d_{\lc_{(\al,\Om,m)}^{-1}} \mu(\vr) \Rightarrow \exp\{-V_\Om(\phi)\} \le 1 . \tag{11}
\end{equation}

Since $(d_{\lc_{(\al,\Om,m)}^{-1}} \mu(\phi))$ is a truly functional probability measure for $\al>\frac D2$, an application of the Lebesgue dominated theorem gives the finitude of the associated generating functional eq(8) for functional interactions of the form eq(10). \qed

\vskip .2in

Note that remains a non trivial problem to evaluate the $n$-point field correlation functions on this proposed scheme of ours. However the same reasoning below can be applied to prove that the exponential regularized $n$-point functions are finite. Namelly
\begin{align*}
\, &\int_{L^2(\Om)} (d_{\lc_{\al,\Om,m}^{-1}} \mu(\phi)) \exp \{ -V_\Om(\phi) \} \\
&\qquad \times \exp \left( i \int_\Om (e^{-\delta\phi^2} \phi)(x) j(x) d^D x \right) \\
& \qquad = Z_{(\al,\Om,m^2,\delta)} (j(x)) \tag{12}
\end{align*}
is an analytical functional on $L^2(\Om)$. \qed

\vskip .2in

\noindent
{\bf Theorem 3.} {\it The analytic regularized, finite volume generating functional $Z_{\al,\Om,m}[j]$ is defined now on the Sobolev Space $H_0^m(\Om)$ (now $\Om$ denoting an open set with compact closure) if $\al$ is sufficiently higher $(2\al-2m-D>0)$ and for Drichlet conditions on the field configurations on the $\Om$-boundary $(\phi|_{\po\Om}\equiv 0)$.
}

\vskip .2in

On the light of such theorem, one can use the Sobolev immersion Theorem $H_0^m(\Om)\subset C^P(\Om)$, for $P<m-D/2$, to have continuous or even differentiable euclidean field sample configurations on the theory's path integral for higher order free field path integral.

At this point appears the very difficult problem of cut-off remotions $\text{vol}(\Om)\to\infty$, $\al\to 1$, $\delta^2\to 0$ or and $m^2\to 0$ in these analitically ``regularized''\, field thories. In next section one implement the finite volume and the analitically regularized remotion on a class of non trivial massless scalar field theories, just for exemplifying that cutt-off remotions on ours proposed regularized path integrals are as possible as well.

Let us point out that theorem 3 can be considered as a sort of generalized Wiener theorem on the continuity of Brownion motions for ours volume-analitically regularized euclidean fields path integrals.

The proof of Theorem 3 is again a direct result of the Minlos-Bochner theorem ([1], [2], [6]).
\begin{equation}
\int_{L^2(R^D)} d_{\lc_{\al,\Om,m}^{-1}} \mu(\phi) ||\vr^2||_{H^m(\Om)} < \infty \tag{13-a}
\end{equation}
if
\begin{equation}
Tr_{H^m(\Om)} [\lc_{\al,\Om,m}^{-1}] < \infty. \tag{13-b}
\end{equation}

This can be verified by a direct computation
\begin{align*}
\, & Tr_{H^m(\Om)} [\lc_{\al,\Om,m}^{-1} ] \\
&\qquad = \frac1{\text{vol}(\Om)} \left\{ \int_{R^D} d^D k \int_{R^D} d^Du \frac{k^{2m}}{u^{2\al}+m^2} |\hat I_\Om(k-U)|^2 \right\} < \infty \tag{13-c}
\end{align*}
if $m+D<\al$.

The reader can check eq(13-c) by means of the finitude condition of the integral \footnote{See also Appendix B for a discussion on the sample differentiability of the Euclidean Field Theory Path Integraly on Hilbert Spaces.}
\begin{equation}
\int d^Dk \frac{K^{2m}}{K^{2\al}+m^2} < \infty \quad \text{ if } \quad 2\al-2m-2D>0.\tag{13-d}
\end{equation}

\section{On the cut-off remotion on a two-dimensional Euclidean QFT model}

We start our studies in this section by considering the (bare) euclidean functional integral on a finite volume smooth compact region $\Om\subset R^2$ as given below (see ref.[2]) for $\al>1$, associated to a real scalar field on $\re^2$
\begin{align*}
Z_\al(j(x)) &= \frac1{Z_\al(0)}  \left\{ \int_{L^2(\Om)} (d_{\lc_\al^{-1}} \mu)(\vr) \right\} \\
&\quad \times \exp \left(-g_{\text{bare}} \int_{\re^2} V(\vr(x))d^2 x \right) \\
&\quad \times \exp \left( i \int_{\re^2} j(x) \vr(x) d^2 x \right). \tag{14}
\end{align*}

Here the functional measure on the path space of real square integrable function on $\Om$, denoted by $L^2(\Om)$ is given through the Minlos Theorem (see Section 1) for real field sources $j(x)$ on $L^2(R^2)$ and the parameter $\al$ on the range $\al>1$. Here $m^2$ is a mass parameter eventually vanishing at the end of our estimate, since we are only interested on the ultraviolet field singularities
\begin{align*}
\, & \exp \left\{ - \frac12 \int_{\re^2\times\re^2} d^2xd^2y \,\, j(x)(\chi_\Om(x)((-\Delta)^{-\al}+m^2)(x,y) \chi_\Om(y)) j(y) \right\} \\
&\quad = \int_{L^2(\Om)} (d_{\lc_\al^{-1}} \mu)(\vr) \exp( i \langle j,\vr\rangle_{L^2(R^2)} ). \tag{15}
\end{align*}

The interaction potential is a continuous function vanishing at infinite such that it posseses an essential bounded $L^1(R)$ Fourier Transform (for instance $V(x)=\frac{\la}{4!} e^{-\delta x^2} x^4,$ etc...). In others words:
\begin{equation}
|| V||_{L^\infty(\Om)} \le ||\widetilde V||_{L^\infty(R)} <\infty. \tag{16}
\end{equation}

Let us show that by defining the bare coupling constant by the renormalization prescriptions $(v=\text{vol}(\Om))$
\begin{equation}
g_{\text{bare}}(\al,v) = \frac{g_{\text{rem}}}{(1-\al)^{1/2}v} \tag{17}
\end{equation}
The functional integral eq(1) has a finite limite at $\al\to1$, $v\to\infty$, when understood in the R.P. Feynman sense as an expansion perturbative on the renormalized constant $g_{\text{rem}}$. Namely (see Chapter 5, \S 5.2, eq(5.11), [7])
\begin{equation}
Z_{\al=1}(j(x)) = \lim_{V\to\infty} \left( \lim_{N\to\infty} \left( \lim_{\al\to 1} \left( \lim_{m^2\to0} I_N (g_{\text{bare}}(\al,v), [j])
\right) \right) \right).\tag{18}
\end{equation}

\vskip .2in

\noindent
{\bf Theorem 1.} {\it The functional $I_N(g_{\text{bare}}(\al,v),[j])$ satisfies the upper bound at the limit $\al\to1$
\begin{align*}
\, & \lim_{N\to\infty} \left( \lim_{\al\to1} |I_{N,\al}(g_{\text{bare}}(\al,v),[j]|) \right) \\
&\qquad \le C^2 \exp(C) \tag{19-a}
\end{align*}
where the constant $C$ is given by
\begin{equation}
C=(4\pi)^{1/2} g_{\text{ren}} ||\widetilde V||_{L^\infty(R)}.\tag{19-b}
\end{equation}
}
\vskip .2in

\noindent
{\bf Proof:} By noting that $\int_{C(\Om)}(d_{\lc_\al^{-1}}\mu)(\vr)\exp(i k\vr(x))=0$, one has the following result eq(8) (see [2]), where the integral kernel of the our ``free'' propagator is given explicitly by (for $\al>1$; see Appendix A)
$$
\lc_{\al,m^2=0}^{-1}(x_i,x_j)=\chi_\Om(x) \left[ \left( \frac1{(2\pi)}(|x_i-x_j|^{2(\al-1)}) \left( \frac{\Ga(1-\al)}{\Ga(\al)} 2^{2(1-\al)} \right) \right) \right] \chi_\Om(y).
$$
\begin{align*}
\, & I_{N,\al}(g_{\text{bare}}(\al,v),[j]) \\
&\quad = \sum_{n=0}^N \Big\{ \frac{(-1)^n}{n!}(g_{\text{bare}}(\al,v))^n \int_\Om d^2 x_1\dots d^2 x_n \\
&\qquad \times \int_R \frac{dk_1}{(2\pi)^{1/2}} \cdots \int_R \frac{dk_n}{(2\pi)^{1/2}} (\widetilde V(k_1)\cdots \widetilde V(k_n)) \\
&\qquad \times \Big[ \int_{C(\Om)}(d_{\lc_\al^{-1}}\mu)(\vr)\exp \left( \sum_{\ell=1}^n ik_\ell \vr(x_\ell) \right) \\
&\qquad \times \exp \left( i \int_\Om j(x)\vr(x) d^2 x \right) \Big] \Big\} \tag{20}
\end{align*}

As a consequence of the positiviness of the ``kinetic''\, Green function $\lc_\al^{-1}(x,y)$; one has
\begin{align*}
\, & (I_{N,\al} (g_{\text{bare}}(\al,v),[j]) \le 1 + \\
&\quad + \Big \{ \sum_{n=1}^N \frac{(|(g_{\text{bare}}(\al,v))|)^n}{n!} (||\widetilde V||_{L^\infty(\Om)})^n \int_\Om d^2 x_1\dots d^2 x_n \\
&\quad \times \Big[ \det_{N\times N}^{-1/2} [\lc_{\al,m^2}^{-1}(x_i,x_j)] \Big] \Big\}. \tag{21}
\end{align*}

Note that due to the continuity on the infrared cut-off mass parameter, it is possible to consider directly its limit on the determinant formed by the Green's functions.

At this point we note the Taylor expansion of the below written object
\begin{align*}
L_N(\al,v) &= \int_\Om d^2x_1\cdots d^2x_n \det_{N\times N_{\substack{1\le i\le N \\ 1\le j\le n}}}^{-1/2} [\lc_\al^{-1}(x_i,x_j)] \\
&= (1-\al)^{N/2} C_n+(1-\al)^{\frac N2+m_1} C_{N+1}+\dots \tag{22}
\end{align*}
with
\begin{equation}
C_n=v^n (4\pi)^{N/2} \left( \det_{\substack{1\le i\le N \\ 1\le j\le N}} [A_{ij}]\right)^{-1/2} \tag{23-a}
\end{equation}
and the matrix $[A_{i,j}]$ is defined by the rule
\begin{equation}
[A[_{ij} = \begin{cases}
0 \quad \text{ if } \quad i=j \\
1 \quad \text{ if } \quad i\ne j \end{cases}
\tag{23-b}
\end{equation}

It yield thus ([2], [3]) for $N>1$
\begin{align*}
\, &\lim_{V\to\infty} \Big( \lim_{\al\to1} |(g_{\text{bare}}(\al,v)L_N(\al,v))| \\
&\quad = \lim_{V\to\infty} \Big \{ \lim_{\al\to 1} \left[ \left| \left( \frac{g_{\text{ren}}}{(1-\al)v} \right)^N L_N(\al,v) \right| \right] \\
&\quad = \lim_{V\to\infty} \Big( \lim_{\al\to1} \Big \{ \left( \frac{|(g_{\text{ren}})|^N}{v^N |(1-\al)|^{N/2}} \right) \\
&\qquad \times \Big[ |(1-\al)|^{N/2} V^N(4\pi)^{N/2} (|(-1)(N-1)(-1)^N|)^{-1/2} \Big] \Big \} \Big) \\
&\quad = (|g_{\text{ren}}|)^N \frac1{(N-1)^{1/2}} ((4\pi)^{1/2})^N \Big) \tag{24}
\end{align*}

We have thus the uniform bound on the ``interaction order'' $N$ in our Euclidean QFT model
\begin{align*}
\, & \lim_{V\to\infty}\lim_{\al\to1} 
(|I_{N,\al}(g_{\text{bare}}(\al,v),[j])|) \\
&\le 1+ \Big \{ \sum_{N=2}^M \frac{(4\pi)^{N/2}}{N!} \frac{(|g_{\text{ren}}|)^N}{(N-1)^{1/2}} \\
&\quad \times [||\widetilde V||_{L^\infty(\Om)}]^N \Big \} \\
&\le C^2 \left( \sum_{N=0}^\infty \frac{C^N}{N!} \right) \\
&= C^2 \exp (C) \tag{25}
\end{align*}
with
\begin{equation}
C=|g_{\text{ren}}| \cdot ||\widetilde V||_{L^\infty} (4\pi)^{1/2}. \tag{26}
\end{equation}

Note that we have used the elementary estimate to arrive at eq(25) for $N>1$
\begin{equation}
\frac1{(N-1)^{1/2}}\le 1, \quad \text{ for } N\ge2. \tag{27}
\end{equation}

We conclude this, that the functional path integral eq(14) under the renormalization coupling constant eq(17) and rigorous Feynman perturbative definition eq(18) has a finite limit for $\al=1$.

It is worth that one could also consider the most general multiplicative renormalization including the functional form of the interaction
\begin{equation}
g_{\text{bare}}(\al,v||\widetilde V||_{L^\infty(\Om)})= \frac{g_{\text{ren}}}{(1-\al)^{1/2} v ||\widetilde V||_{L^\infty(\Om)}}. \tag{28}
\end{equation}

Now allowing interactions satisfying the constraint $\widetilde V(k)=\lim\limits_{\ell\to\infty}\widetilde V_\ell(k)$ with $||\widetilde V_\ell(k)||_{L^\infty(\Om)} = \ell \in \nb^+$.

It is worth to recall that we have proven that the full generating functional eq(14) at $\al\to1$ as defined by a Feynman's perturbative series: Feynman's diagrammas renormalized order by order in a power serie expansion on the bare coupling constant is finite and it is a continuous functional on the source space $j(x)\in L^2(\Om)$.

However it appear that the use of the propagator prescription
\begin{align*}
\, & \hat{\lc}_{\al,m^2=0}^{-1} (x_i,x_j) = \chi_\Om(x) \Big( \Big[ \frac1{4\pi}\, \frac{\Ga(1-\al)}{\Ga(\al)} 2^{2(1-\al)} \\
&\quad \times |x-y|^{2(\al-1)} \Big] - \frac{4\pi}{(1-\al)} \Big) \chi_\Om(y), \tag{29}
\end{align*}
which converges on the $D'(R^2)$ L. Schwartz distributional sense to the usual non positive definite two-dimensional Laplacean Green function for $\al\to1$ does not lead to well defined Euclidean QFT generating functional. A result already expected since Massless $2D$ Euclidean Q.F.T. Theories built already as perturbation around free scalar Massless fields on $R^2$ do not make mathematical sense due to the fact that the two-dimensional Laplacean Green function does not belongs to the ``Fourier Transformable'' Tempered Distributional Space $S'(R^2)$, a fact already observed a long time ago by S. Coleman ([4]) and fully used by G. Hoft on his studies on $(QCD)_2$ -- solubility at large number colors ([5]).

Another point worth call attention in this Section is that the same proof works out for a class of four-dimensional analitically regularized Euclidean Field theories with the ``Free kinetic operator'' defined through the Minlos's theorem on a finite volume region $\Om\subset R^4$
\begin{align*}
\, & \exp \Big\{ - \frac12 \int_{R^4} d^4 x\int_{R^4} d^4 y \\
&\qquad \times j(x) \Big\{ \chi_\Om(x) \Big[ (-\Delta^2)^{-\al} +m_0^2 \Big] \chi_\Om(y) \Big\} j(y) \\
&\quad = \int d_{\lc_{\al,m_0^2}^{-1}}\mu(\vr) \exp \Big( i \int_{R^4} \vr(x)  j(x) d^4x \Big)\Big\}. \tag{30-a}
\end{align*}

Here the Integral Kernel of the square $D$-fimensional Laplacean is given by (for $\al>1$)
\begin{equation}
(-\Delta^2)^{-\al} = \frac{\Ga(\frac D2-2\al)}{\Ga(2\al)2^{ 4\al} \pi^{D/2}} (|x-y|^{4\al-D}). \tag{30-b}
\end{equation}

Finally, we call attention that into another publication we will address to the ``differentiability'' of the generating functional eq(1) at $\al\to1$ as defined in the Bulk of this section. However it is straithforward to obtain such differentiability for sources $j(x)$ coupled to field configurations interaction of the form $\exp(-\delta\vr^2(x)) \vr(x)$. Note that in this case, the $N$-point Taylor's coeficients of $Z[j(x)]$ are explicitly given by
\begin{align*}
\, & \frac{\delta^N Z[j(x)]}{\delta j(x_1) \cdots \delta j(x_N)}\Big|_{j(x)\equiv 0}\\
&\quad = \int_{C(\Om)} (d_{\lc_\al^{-1}}\mu)(\vr) \left( \prod_{\ell=1}^N [\exp(-\delta \vr^2(x_\ell))\vr(x_\ell)] \right) \\
&\quad < \infty, \tag{31}
\end{align*}
since the domain of the above functional integral
for $\al>1$ is the space of measurable square integrable functions on $\Om$ (and for $\delta>0$)
\begin{equation}
||e^{-\delta\vr^2(x)} \vr(x)||_{L^\infty(R)} = (\max_{x\in R} |e^{-\delta x^2}x|) =C<\infty, \tag{32}
\end{equation}
leading to the finitiness of eq(31) by the use of the Lebesgue dominated convergence theorem.

The limite of $\delta\to0$ on the momentums eq(31) will appears elsewhere.

Finally we wishe to point out that non trivial homological topology of the compact planar two-dimensional domain $\Om$ ([8]) in ours path integral can be easily taken into account by the $\Om$ set indicator function $\chi_\Om(\Om)$ on eq(30-a) of this section, specially on Fourier Space by means of the $\Om$-domain Fourier Integral form factor for $\Om$ with holes inside
\begin{equation}
\hat I_k(\Om)=\left( \int_\Om d^2 \xi \exp(i k \xi) \right); \tag{33}
\end{equation}
which appears on the expression of the theory's propagator on momentum space for general $R^D$ space-time
\begin{equation}
\hat{\lc}^{-1} (k,k')=\int_{R^D} \left( \frac{\hat I_\Om(k-p)\hat I_\Om(p-k')}{p^{2\al}+m^2} \right) d^D p \tag{34}
\end{equation}
and leading thus to the Feynman diagrammotic generating functional on the Fourier Space
\begin{align*}
\, & Z[\tilde j(k)]/Z[0] \\
&\quad = \exp \left\{ - \int_\Om d^Dx \left(V \left( \frac1{(2\pi)^{D/2}} \int_{R^D} d^D k e^{+ikx} \left( \frac\delta{\delta \tilde j(k)} \right) \right) \right) \right\} \\
&\qquad \times \exp \Big\{ - \frac 12 \int_{R^{2D}} dk dk' \tilde j(k) \\
&\qquad \times \left( \int_{R^D} dp \frac{\hat I(k-p)\hat I(p-k')}{p^{2\al}+m^2} \right) \tilde j(k') \Big \} \tag{35}
\end{align*}

As a last remark, we conjecture that the ultra-violet limit $\al\to1$ on the usual correlations functions associated to our path integral should expected to be finite. The argument follows by considering $||\widetilde V||_{L^\infty(\Om)}=1$, since $x\to0$ and thus, obtaining the general structure of the (for instance) two-point function at perturbative order $N$
\begin{align*}
 \langle\vr(x_1)\vr(x_2)\rangle & \overset{\al\to 1}{\sim} - (\lc_{\al,m^2=0}^{-1}(x_1,x_2)) \\
&\quad +  \sum_{p=1,q=1}^N \Big\{ \lc_{\al,m^2=0}^{-1}(x_1,x_p) \\
&\qquad \times [\lc_\al^{-1}(x_i,x_j)]_{pq}^{-1} \lc_{\al,m^2=0}^{-1}(x_q,x_2) \Big\} \tag{36}
\end{align*}
and noting the Laplace formula for evaluate the inverse of the propagator matrix
\begin{align*}
\, & [\lc_\al^{-1} (x_i,x_j)]_{pq}^{-1} = \frac1{\det_{N\times N}[\lc_\al^{-1}(x_i,x_j)]} \\
&\qquad \times (\cb(x_i,x_j)]_{qp} ,\tag{37}
\end{align*}
with the cofactor matrix $[\cb(x_i,x_j)]_{qp}$ associated to the $[\lc_\al^{-1}(x_i,x_j)]$ propagator matrix eq(21). One expects thus that the singular behavior for $\al\to 1$ of the determinant
$$
\det_{N\times N}[\lc_\al^{-1}(x_i,x_j)] \overset{\al\to1}{\sim} (1-\al)^{-N},
$$
cancels out with the factor
$$
\sum_{p=1,\ve=1}^N \lc_{\al,m^2}^{-1}(x_1,x_p)
[\cb(x_i,x_j)]_{qp} \lc_{\al,m^2=0}^{-1} (x_q,x_2) \sim (1-\al)^{-N},
$$
on eq(36).

\section{On the construction of the Wiener Measure}

On next Section 4, we intend to analyze the somewhat different functional-path integral on functional space, mainly due R.P. Feynmann and M. Kac: the so called geometrodynamical end points fixed field propagator.

However, such objects to be defined mathemat\textsl{}ically, one must review the construction of the famous Wiener path measure ([6]--[7]). This is our objective in this short section.

Let us first introduce some notations and mathematical objects.

We first write the time fixed Heat equation green function on the one-point compactified of the real line $R$, the interval $[-\frac\pi2,\frac\pi2]$ as a integral kernel of a continuous linear functional on the compact support continuous function $f$ on $R$. So, let $\ve>0$ and $f\in C_c(R)$
\begin{align*}
L_\ve(f) &= \int_R \left( \frac{\exp(-|x-y|^2/2\ve)}{(2\pi\ve)^{1/2}} \right) f(y) dy \\
&= \int_{-\frac\pi2}^{\frac\pi2} 
\overbrace{\left( \frac{e^{-|tg(\te_x)-tg(\te_y)|^2/2\ve}}{(2\pi\ve)^{1/2}\cos^2\te} \right)}^{:=\hat G_0(x,y,\ve)} f(tg\te) d\te . \tag{38}
\end{align*}

Note that $\supp f(tg\te)\subset(-\frac\pi2,-\frac\pi2)$. For a given $g(\te)\in C_c([-\frac\pi2,\frac\pi2])$, eq(38) defines a positive continuous linear functional on $C_0(\dot R)(f(\infty)=g(-\frac\pi2)=g(\frac\pi2)=0)$.

Let us define the following projective family a positive linear functionals on $C_c(\prod\limits_{n=0}^\infty([-\frac\pi2,\frac\pi2])_n)$,
firstly defined on the $\sigma$-algebra of the infinite variable space $C_c(\prod\limits_{n<0}^\infty \dot R)\equiv C_0((\dot R)^\infty)$. For $N$ a given positive integer fixed, but arbitrary and $\bar x$ a fixed point on $[-\frac\pi2,\frac\pi2]$, we consider the projected positive continuous linear functionals $\left(\ve = \frac1N\right)$;
\begin{align*}
L_{\bar x}^{(N)} (f(x_1,\dots,x_N)) &= \int_{\dot R} dx_1\dots \int_{\dot R} dx_N(f(x_1,\dots,x_N))\Big) \\
&\quad \times (\hat G_0(x,x_1,\ve)\dots\hat G_0(x_{N-1},x_N, \ve)) dx_1\dots dx_N. \tag{39}
\end{align*}

We point out the ``projective'' properties of the family  of positive continuous functionals $\{L_{\bar x}^{(N)}\}$:

\vskip .2in

a) For $M\le N$,
\begin{equation}
L_x^{(M)} (f(x_1,\dots x_M))=L_x^{(N)} (f(x_1,\dots,x_M)) \tag{40}
\end{equation}

\vskip .2in

b) \,\, $C_{c,\text{finite variables}}(\dot R^\infty)=\{ f\in C_c(\dot R^\infty)$, but with finite variables\}  is a dense subset of $C_c(\dot R^\infty)$ (endowed with the usual supremum norm!).

All theses results above remarked show that exists the
$$
\lim_{N\to \infty} L_{\bar x}^{(N)} = L_{\bar x}^{(\infty)} \quad \text{ on }\,\, C_c(\dot R^\infty).
$$

Since $\dot R^\infty$ is a compact topological space one can apply the Riesz Markov theorem to represent $L_{\bar x}^\infty$ through a well defined measure on $\dot R^\infty$ (the Bare $\sigma$-albegra of $\dot R^\infty$).

Note that one could take $\ve=t/N$ with $t>0$, a real fixed, and this obtain the famous Wiener measure ending at $\bar x$ at time $t$
\begin{equation}
L_{(\bar x,t)}^\infty(f)=\int_{\dot R^\infty} d_{(\bar x,t)}^{\text{Wiener}} \hat\mu[g(\sigma)] f(g(\sigma))\tag{41}
\end{equation}
where $g(\sigma)\in\dot R^\infty$ is identified with the set of all real functions on $\dot R$ [the ``compactified'' Wiener path trajectory], with the domain $\sigma\in[0,t]$.

It is worth to remark that on eq(39), all the ``time parameters'' are at the same value $t=\ve$.

It is an open problem to show the existence and unicity of the Wiener measure $d_{(\bar x,t)} \mu[g(\sigma)]$ under general (different) time steps on eq(39).

It is worth also to note that $f\in C_c(\dot R^\infty,R)$ by the hypothesis of the Riesz-Markos theorem ([7]).

At this point if is argued that there is a unique ``pull-back'' of the above constructed Wiener measure on the space of compact paths to the full $R$ paths. Namely, for $F\in C_c(C(R,R),R)$ and $x\in R$
\begin{equation}
L_{(x,t)}^\infty(F) =\int d_{(x,t)}^{\text{Wiener}} \mu(X(\sigma))F(X(\sigma)).\tag{42}
\end{equation}

The above construction generalizes straightforwardly for $R^D$ $(D >1)$.

We have thus the following theorem (Feynmann Wiener-Kac): Let $(-\Delta)$ be the essential self-adjoint extension of the usual Laplacean acting  on $C_c^\infty(R^D)$.

We have the formula ([7]), for $F\in C_c(R^D)$ on the sense the topology of $C_c(R^D)$
\begin{align*}
\, & (e^{-\frac t2\Delta} F)(x) \\
&\quad = \int (d_{(x,t)}^{\text{Wiener}} \mu(X(\sigma))F(X(t)). \tag{43}
\end{align*}

For general $F\in L^2(R^D)$, eq(43) is obtained by (unique) extension, since $((\ov{C_c(R^D)})_{L^2(R^D)}=L^2(R^D))$.

It is important to call attention that due to the $C^\infty$-regularizing property of the heat Kernel eq(38), the functional integral representation eq(43) remains correct in the $L^2(R^D)$ sense for $F(w)=\delta^{(D)}(w-y)$ and leading thus to the formal Brownian Bridge path integral measure representation for the Heat Kernel
\begin{equation}
\langle x|e^{-\frac t2\Delta}F|y\rangle \overset{L^2(R^D)}{=} \int_{X(0)=y}^{X(t)=y} d^{\text{Wiener}} \mu(X(\sigma)) F(X(t)) \tag{44-a}
\end{equation}
or in the correct mathematical meaning of the above written eq(44-a) for $f$ and $g\in L^2(R^D)$
\begin{equation}
\int_{R^D} f(x) \langle x|e^{-\frac t2\Delta}F|y\rangle \bar g(y)= \int_{X(0)=y}^{X(t)=y} d^{\text{Wiener}} \mu(X(\sigma)) (f(X(t))\bar g(X(0))). \tag{44-b}
\end{equation}

\section{On the Geometrodynamical Path Integral}

Sometimes it appears to be useful for calculational purposes on euclidean quantum field theory to give a generalized meaning for the Brownian Bridge Wiener path integral eqs(44-a)--(44-b), called now the Geometridynamical propagator connecting a classically observed field configuration $\phi(x,t_1)=\be_{in}(x)$ to another final one $\phi(x,t_2)=\be_{out}(x)$, with $t_2>t_1$.

Let us formulate the problem for the free case of a real scalar field $\phi(x,t)$ with classical action and with Dirichlet boundary conditions on the $D$-dimensional space-time cylinder manifold propagation $D=\Om\times[t_1,t_2]$ with $A$ denoting an inversible positive definite self-adjoint elliptic operator on $\Om$
\begin{equation}
S[D]=\int_D \frac12 \left(\be\left(-\frac{\po^2}{\po t^2}+A\right) \be \right) (x,t) d^{D-1} xdt . \tag{45}
\end{equation}

One wants to give a rigorous mathematical meaning for the Euclidean Feynman-Wheller path integral
\begin{align*}
\, & G[(\be_{in}(x),t_1); (\be_{out}(x),t_2)] \\
&\qquad = \int_{\be(y,t_1)=\be_{in}(x)}^{\be(x,t_2)=\be_{out}(x)} D^F[\be(x,t)]\exp\{-S[D]\}. \tag{46}
\end{align*}

The most usual way to give a mathematical meaning for eq(46) is to use the spectral theorem for $A$ $(A\vr_\mu=\la_\mu\vr_\mu)$ and regard eq(46) as the (enumerable) infinite product of Brownian Bridge Wiener measures eq(44-a) and under the hypothesis that all the field configurations entering on the support of the resulting field path integral measure is of the form
\begin{equation}
\be(x,t)=\sum_{\mu=0}^\infty C_\mu(t)\phi_\mu(s)\in C([t_1,t_2],L^2(\Om)).\tag{47}
\end{equation}

We thus define eq(46) as
\begin{align*}
\, & G[(\be_{in}(x),t_1),(\be_{out}(x),t_2)] \\
&\quad= \prod_{n=0}^\infty \left[ \int_{X_n(t_1)=\be_n^{in}}^{X_n(t_2)=\be_n^{out}} d^{\text{Wiener}} \mu(X_n(\sigma))\times \exp \left( -\frac12 \int_{t_1}^{t_2} (\la_n)^2(X_n(\sigma))^2 \right) \right]. \tag{48}
\end{align*}

Here
\begin{equation}
\be^{in}(x)=\sum_{n=0}^\infty \be_n^{in} \vr_n(x)\tag{49-a}
\end{equation}
\begin{equation}
\be^{out}(x)=\sum_{n=0}^\infty \be_n^{out} \vr_n(x).\tag{49-b}
\end{equation}

Let us note that the enumerable infinite product of Wiener-Harmonic Oscilator measure is still a well behaved $\sigma$-measure on the product measure space $\prod\limits_{n=0}^\infty (C([t_1,t_2],R))_n$. Note that if one user the compactified of the real line as in Section 3, one would gets as the measure space; the compact space
$\prod\limits_{n=0}^\infty C([t_1,t_2],\dot R)_n$.

In the presence of an external source $f(t,x)\in C([0,T],L^2(\Om))$, one has the usual Feynman closed expression in terms of Feynman-Wiener notation for the Wiener-Harmonic oscillator path measures ([7])
\begin{align*}
\, & G[(\be^{in}(x),0); (\be^{out}(x),T),[j(x,t)]]
\\
&= \prod_{n=0}^\infty \int_{X_n(0)=\be_n^{in}}^{X_n(T)=\be_n^{out}} \left\{ D^F[X_n(\sigma)] \exp \left[ -\frac12 \int_0^T \left( X_n \left( -\frac{d^2}{d\sigma^2}+\la_n^2\right) X_k \right)(\sigma) \right] \right\}
\\
&\quad \times \exp \left( \int_0^T d\sigma j_n(\sigma)X_n(\sigma) \right)
\\
&=\prod_{n=0}^\infty \left\{ \sqrt{\frac{\la_n}{Sinh(\la_n T)}} \right.
\\
&\quad \times \exp \left\{ - \frac{\la_n}{2sinh(\la_nT)} \left[(\be_n^{out})^2 \right.\right.
\\
&\quad + (\be_n^{in})^2 cosh(\la_nT)-2 \be_n^{out}\be_n^{in} \Big] \Big\}
\\
&\quad - \frac{2\be_n^{out}}{\la_n}\int_0^T d\sigma j_n(\sigma)sinh(\la_n\sigma)
\\
&\quad - \frac{2\be_n^{in}}{\la_n}\int_0^T d\sigma j_n(\sigma)sinh(\la_n(T-\sigma))
\\
&\quad -\frac2{(\la_n)^2} \int_0^T d\sigma \int_0^T d\sigma' j_n(\sigma) j_n(\sigma')sinh(\la_n(T-\sigma))
\\
&\quad \times sinh(\la_n(\sigma')) \Big\} \tag{50}
\end{align*}

Another more attractive prescription to eq(46), specially useful on String Theory ([7]) is to suppose that the sample space for geometrodynamical propagation is composed of field configurations made by random perturbations of a (fixed) classical field configuration as exposed in \S5.3, eqs(5.41)--(5.47) of ref[7]. However, this method does not appears to be canonically invariant, since all the objects on the theory are dependent of the choosen background field configuration the classical choose field configuration, i.e. for different background field configuration one could obtain different path integrals.

\newpage

\centerline{\huge\bf Appendix A}

\vskip .5in

Let us recall the following integral form of a Fourier Transform of a Tempered distribution $T_f$ defined by a $L^2_{loc}(R^D)$ radial function $f(r)$
\begin{equation}
\fc(T_{f(r)})=T_{\hat F(k)} \tag{A-1}
\end{equation}
with
\begin{equation}
\hat F(k)=(2\pi)^{D/2} \left( \int_0^\infty \frac{f(r) r^{D-1} J_{\frac{D-2}2}(kr)dr}{(kr)^{\frac N2-1}} \right).\tag{A-2}
\end{equation}

By using our proposed distributional sense integral formulae for $\mu$ and $\nu$ complex numbers and $a>0$
\begin{equation}
\int_0^\infty x^\mu J_\nu(ax)dx=2^\mu a^{-\mu-1} \frac{\Ga(\frac12+\frac12 \nu+\frac12 \mu)}{\Ga(\frac12+\frac12\nu-\frac12\mu)}. \tag{A-3}
\end{equation}

One obtains the result on the $S'(R^D)$ sense for $\al\in\cb$
\begin{align*}
\, & \fc\left[ T_{(\frac1{2\pi}\frac{\Ga(1-\al)}{\Ga(\al)} 2^{2(j-\al)} r^{2(\al-1)}} \right] \\
&\qquad = T_{K^{-2\al}}. \tag{A-4}
\end{align*}

The complete distributional sense is given below for $f(x)\in S(R^D)$, with $\hat f(k)=\fc[f(x)]$
\begin{equation}
T_{(\frac1{2\pi}\frac{\Ga(1-\al)}{\Ga(\al)} 2^{2(1-\al)} r^{2(\al-1)}} (f) = T_{k^{-2\al}} (\hat f(k)).\tag{A-5}
\end{equation}

Just for completeness, let us evaluate on the $S'(R^2)$ sense the Fourier Transform below
\begin{align*}
G_\al(x,y,m^2) &=\left( \frac1{\sqrt{2\pi}}\right)^2 \left( \int_{R^2} d^2k e^{ik(x-y)} \frac1{(k^2+m^2)^\al} \right)
\\
&= \frac1{2\pi} \left( \int_0^\infty dk \frac k{(k^2+m^2)^\al} \overbrace{\left( J_0(kr)+J_0(-kr)\right)}^{2J_0(kr)} \right)
\\
&= \frac1{2\pi} r^{2(\al-1)} \times \left(\int_0^\infty dp \frac{pJ_0(p)}{(p^2+m^2r^2)^\al} \right)
\\
&=\frac{(mr)^{1-\al}\cdot K_{1-\al}(mr)}{2^{\al-1}\Ga(\al)} \tag{A-6}
\end{align*}
where we have used the distributional sense integral relation for $\mu$ and $\nu$ complex parameters and $a,b\ge0$:
\begin{equation}
\int_0^\infty \frac{J_V(bx)x^{\nu+1}}{(x^2+a^2)^{\mu+1}}  dx=\frac{a^{\nu-\mu} b^\mu K_{\nu-\mu}(ab)}{2^\mu \mu(\mu+1)} . \tag{A-7}
\end{equation}

Just for completeness, one can use the above exposed formulae to obtain the Integral Kernel of the $S'(R^D)$ distribution $(-\Delta)^{-\al}$. Namely
\begin{equation}
(-\Delta)^{-\al}(x,y)=\frac{\Ga(\frac D2-\al)}{\Ga(\al)2^{2\al}\pi^{D/2} (x-y)^{D-2\al}}.\tag{A-8}
\end{equation}

\newpage

\centerline{\huge\bf Appendix B}

\vskip .5in

\noindent
{\bf Theorem.} {\it Let $A$ be a positive definite trace class operator on $L^2(\Om)$, with spectral resolution $A\vr_n=\la_n\vr_n$ such that $\sum\limits_{n=0}^\infty \la_n n^{2p}=+\infty$ for $p>0$. Let also $d_A\mu(\vr)$ denotes the cylindrical measure asociated to $A$ through the Minlos-Bochner theorem applied to the bilinear source function $Z(j)=\exp\{-\frac12 \langle j, A_j\rangle_{L^2(\Om)}\}$.}

Let $H_p=\{ f\in L^2(\Om) \mid f=\sum c_n \vr_n$ with $\sum\limits_{n=0}^\infty c_n^2 n^{2p}<\infty\}$ be the ``Sobolev'' sequence measurable sub-sets of $L^2(\Om)$. Then we have that for any $p>0$
$$
\mu(H_p)=0.
$$

Roughly this result means that $C^\infty(\Om)$-smooth path integral field sample configurations on the path probability space $(L^2(\Om),d_A\mu(\vr))$ make a set of zero measure Proof ([6]).

By using the following representation of the charachteristic function of the 
sub-set $H_p$
\begin{equation}
\chi_{H_p}(\vr)=\lim_{\al\to0}\left\{ \exp \left[ - \frac\al2\sum_{n=0}^\infty c_n^2 n^{2p} \right] \right\}, \tag{B-1}
\end{equation}
we have the identity $(\vr\in L^2(\Om); \vr \overset{L^2(\Om)}{=} \sum c_n\vr_n )$
\begin{align*}
\mu(H_p) &=\lim_{\al=0}\int_{R^N} dj_1\dots dj_N
\\
&\quad \times \left( \int d_A \mu(\vr)\exp \left( i\sum_{n=0}^N j_n c_n \right) \right)
\\
&\quad \times \exp \left( -\frac1{2\al} \sum_{n=0}^\infty \frac{j_n^2}{n^{2p}} \right)
\\
&\quad \times \left[ \prod_{\mu=1}^N(2\pi\al n^{2p})^{-\frac12} \right]\tag{B-2}
\end{align*}

A firect evaluation of the cylinder path integration on the right-hand side of eq(B-4) give us the following outcome
\begin{align*}
\, &\int d_A\mu(\vr) \exp \left( i \sum_{\mu=0}^\infty j_n c_n \right) \\
&\qquad =  (2\pi)^{N/2} \exp \left\{ -\frac12 \sum_{\mu=0}^\infty \frac{j^2_n}{\la_n} \right\}. \tag{B-3}
\end{align*}

As a consequence we have the final result on the measure of the ``Sobolev Spaces'' $H_p$
\begin{align*}
\mu(H_p) &= \lim_{\al\to0} \lim_{N\to\infty} \Big \{ (2\pi^{N/2} \left( \prod_{\mu=1}^N(2\pi\al\mu^{2p})^{-\frac12} \right) \\
&\quad \times \left( \prod_{n=1}^N \left( \la_n+\frac1{\al n^{2p}} \right)^{-\frac12} \right) \Big\} \\
&\le \lim_{\al\to0} \left\{ \lim_{N\to\infty} \exp \left(-\frac12\al \sum_{\mu=1}^N \la_n n^{2p} \right) \right\} \\
&= \begin{cases}
0 \quad \text{ if } p>0 \\
1 \quad \text{ if } p=0 
\end{cases} \quad \text{(Minlos-Bochner Theorem)} \tag{B-4}
\end{align*}
where we have used the straightforward  identity
$$
\prod^N \left( \frac1{1+\al_n}\right)^{1/2} \le \left( \frac 1{1+\sum^N \al_n} \right)^{1/2} \le \exp \left\{ -\frac12 \sum^N \al_N \right\}.
$$
\qed 

\vskip .2in

As it is usual to expect that $C^\infty(\Om)\subset \bigcap\limits_{p=0}^\infty H^p(\Om)$ (where $H^p(\Om)$ denotes the usual function Sobolev Spaces on $\Om$), the theorem of this appendix as expressing the fact that differentiable sample on non enough sufficiently regularized euclidean path integrals makes a set of zero measure. And classical smooth field
configurations being useful only in the realm $\mu$ of formal saddle-point (WKB) path integral evaluations. So one can not manipulate path integral integrands with College Calculus rules.

For the less restringent condition of path integral sample continuity, one has to use our generalization of the Wiener theorem eq(9).

In the general case of non-Gaussian cylindrical measures, one should imposes the bound restriction below, as a reasonable thechnical condition
\begin{equation}
\sup|Z(j)| \le C\left[ \exp \left \{ -\frac12 \langle j,Aj\rangle\right\} \right] \tag{B-5}
\end{equation} 
for some positive definite trace class positive operator $A\in\oint_1^+(L^2(\Om))$ and $C>0$, in order to obtain the validity of our theorem -- Appendix B.

\newpage

\centerline{\huge\bf Appendix C}

\vskip .5in

In this somewhat pedagogical appendix, we intend to presente a formal operational functional calculus to write the cylindrical Fourier Transforms as an inversible operation. We present such formal results in order to highlight the necessity of a clean distribution theory in Hilbert Spaces, get to be developed ([7]).

So let us consider a trace class, inversible and strictly positive operator $A^{-1}$ acting on a separable Hilbert Space $H$.

Let $f(x)\in L^1(d_A\mu(x)(x),H)\cap L^\infty(d_A\mu(x),H)$. Since the function $\exp i\langle x,k\rangle$, for a $k\in H$ fixed is bounded, the $L^1$-Hilbert Space Fourier Transform is well defined
\begin{equation}
\hat F(k)=\int_H f(x)\exp i\langle x,k\rangle d_A\mu(x).\tag{C-1}
\end{equation}

In the Physicist's operational notation for the cylindrical measure with $D^F[x]$ denoting the Feynman Formal (when $H$ realized as some $L^2(\Om)$) 
\begin{equation}
 d_A\mu(x)=\det^{+\frac12}(A) \exp(-\frac12(x,Ax)_H) D^F[x] \tag{C-2}
\end{equation}

It still to be an open problem in Analysis in Infinite Dimensions or and Hida calculus to obtain an rigorous mathematical inversion formula for eq(C-1). However, it is fully possible to write an inversion formula for eq(C-1) in a more larger vectorial space: the called algebraic dual space of $H$ i.e: $H^{alg}$. Let us sketchy for completeness such result of ours. As a first step one introduces an one-parameter $\ve$ $(\ve\in[0,1])$ family of auxiliary inversible operators $\ve^{-2}\cb(\ve)$ on $\lc_{\text{bounded}}(H,H)$ such that $\cb(0)=\bold 1$. Let us now consider the new (perturbed) family of operator $B(\ve)=(A+\ve^{-2}\cb(\ve))^{-1}$. Note that $B(\ve)$ exists for $\ve$ small enough and $B^{-1}(\ve)=A+\ve^2(\cb(\ve))$ do not belong to the trace class althought being positive definite. But even in such situation one can define a cylindrical measure in the more ample space $H^{alg}$ through the positive-definite charachteristic functional associated to the operator $B^{-1}(\ve)$
\begin{equation}
Z_B[x]=\int_{H^{alg}} d_{B(\ve)}\nu(X) e^{i X(x)}=\exp \left\{ -\frac12 \langle x,(B(\ve))^{-1} x\rangle_H \right\}.\tag{C-3}
\end{equation}

Let us choose our interpolating family of bounded, strictly positive operators $\cb(\ve)$ such that for $\ve>0$, $\det^{-\frac12}(\cb(\ve))<\infty$. We now define the following (continuous) linear functional on $H$ for each $\ve>0$, under the non-proved hypothesis that $H$ is a sub-set of $H^{alg}$ of non zero measure
\begin{equation}
I^{(\ve)}(x,[\hat F])=\det^{-\frac12} (\frac1{\ve^2} \cb(\ve)) \left\{ \int_{H^{alg}\cap H} \hat F(k) e^{-ik(x)} d_{B(\ve)} \nu (k) \right\}.\tag{C-4}
\end{equation}

Now if one substitutes eq(C-4) into eq(C-1) and by applying the Fubbini  theorem to the Product Measure Space $(H\times H^{alg}; d_A\mu \otimes d_{B(\ve)}\nu)$, one obtains the result
\begin{align*}
I^{(\ve)}(x) &= \det^{-\frac12} (\frac1{\ve^2} \cb(\ve))
\\
&\quad \times \Big\{ \int_{H^{alg}} \left[ \int_H f(z) e^{i\langle z,k\rangle} d_A\mu(z) \right] e^{-i(x,k)} d_{B(\ve)} \nu(k)= \det^{-\frac12}(\ve^{-2}\cb(\ve))
\\
&\quad \times \left\{ \int_{H^{alg}}\int_H f(z) e^{i\langle z-x,k\rangle} (d_A \mu(z) d_{B(\ve)} \nu(k)) \right\}
\\
&= \det^{-\frac12} (\ve^{-2}(I(\ve))) \times \left\{ \int_H f(z) Z_{B(\ve)}(z-x) d_A\mu(z) \right\}
\\
&=\int_H f(z) e^{-\frac12(z-x,A(z-x))_H}
\\
&\quad \times \left[ \det^{-\frac12}(\ve^{-2}\cb(\ve))\exp \left( -\frac1{2\ve^2} \langle z-x; \cb(\ve)(z-x)\rangle \right)_H \right] d_A\mu(z). \tag{C-5}
\end{align*}

At this point we take from the Hida Calculus, the formal definition of the Dirac delta functional on Hilbert Spaces
\begin{align*}
\, & \lim_{\ve\to0^+} \det^{-\frac12} (\ve^{-2}\cb(\ve)) \exp \left[ -\frac1{2\ve^2} \langle(z-x),\cb(\ve)|z-x|\rangle_H \right] \\
&\qquad = \delta_H^{(F)} (z-x)\quad \text{ on }\quad S'(H). \tag{C-6}
\end{align*}

As a consequence, one has the operational result
\begin{align*}
I(x) &:= \lim_{\ve\to0^+} I^{(\ve)} (x)
\\
&=\int_H f(z) \exp \left( -\frac12 \langle z-x,A(z-x)\rangle_H \right)
\\
&\quad\times \delta_H^{(F)} (z-x) d_A\mu(z)
\\
&:= f(x) e^{-\frac12\langle x,Ax\rangle} \det^{+\frac12}(A). \tag{C-7}
\end{align*}

As a consequence we have the operational formulae for Fourier Transforms in separable Hilbert Spaces.

If one has the usual cylindrical Hilbert Space Transform
\begin{equation}
\hat F(k) := \int_H f(x) \exp(i\langle x,k\rangle_H) d_A\mu(x) \tag{C-8}
\end{equation}
then formally, one has the ``inversion'' formula on the algebraic dual of $H$
\begin{align*}
f(x) &= e^{+\frac12\langle x,Ax\rangle}\det^{-\frac12}(A) \\
&\quad \times \lim_{\ve>0} I^{(\ve)}(x,[\hat F]).\tag{C-9}
\end{align*}

Anyway the necessity of using mathematically rigorous infinite-dimensional Fourier Transforms on Tempered Schwartz Distributions has not appeared fully yet on mathematical physics metrods. However, on light of the results presented in this paper, the time for such endoavours may be approaching ([7]).

\vskip .3in

\noindent
{\bf Acknowledgments:} Thanks to professor D. Pickrell of Mathematics Department of University of Arizona for discussions on $P(\phi)_2$ Field Theories on Riemman Surfaces (ref[8]).

\vskip .5in

\noindent
{\bf REFERENCES}

\vskip .2in

\begin{itemize}

\item[{[1]}] J. Glimm and A. Jaffe, Quantum Physics, Springer, New Yorkl, NY, USA, $2^{nd}$ edition, 1987.

- B. Simon, The $P(\phi)_2$ Euclidean (Quantum) Field Theory, Princeton University Press, Princeton, NJ, USA, 1974.

\item[{[2]}] Luiz C.L. Botelho, Some Comments onj Rigorous Finite-Volume Euclidean Quantum Field Path Integrals in the Analytical Regularization Scheme -- Hindawi Publishing Corporation, Advances in Mathematical Physics, vol 2011, Article ID 257916, 

DOI: 10.1155/2011/257916.

\item[{[3]}] Luiz C.L. Botelho, ``A simple renormalization scheme in random surface theory'', Modern Physics Letters B, vol 13, No. 6--7, pp. 203--207, 1999.

\item[{[4]}] Green, M.R., Schwarz, J.L., Witten, E., Superstring Theory, Cambridge Monographs on Mathematical Physics, vol 182, CUP, Cambridge (1996).

\item[{[5]}] B. Klaiber, in Lectures in Theoretical Physics: Quantum Theory and Statistical Theory, edited by A. O. Barut.  Gordon and Breach, New York, 1960, vol XA, pp. 141--176.

\item[{[6]}] Luiz C.L. Botelho, A note an Feynman Kac path integral representations for scalar wave motions, Random Operators and Stochastic Equations (print), v. 21, pp. 271--292, (2013).

- Luiz C.L. Botelho, Semi-linear diffusion in $R^D$ and in Hilbert Spaces, a Feynman-Wiener path integral study, Random Oper. Stoch. Equ-19 (2011), Issue 4, pages 361--386, DOI 10.1515/Rose 2011.020.

- Luiz C.L. Botelho, A method of integration for wave equation and some applications to wave physics, Random Oper. Stoch. Equ-18 (2010), No. 4, 301--325.

- Luiz C.L. Botelho, Non-Linear Diffusion and Wave Damped Propagation: Weak Solutions and Statistical Turbulence Behavior, Journal of Advanced Mathematics and Applications, vol 3, 1--11, (2014).

\item[{[7]}] Luiz C.L. Botelho, Lecture Notes in Applied Differential Equations of Mathematical Physics World Scientific, (2008), Singapore ISBN: 10981-281-457-4.

\item[{[8]}] Pickrell, Doug, $P(\phi)_2$ Quantum Field Theories and Segal's Axioms. Commun. Math. Phys. 280, 403--425, (2008).

\item[{[9]}] Luiz C.L. Botelho, On the rigorous ergodic theorem for a class of non-linear Klein Gordon wave propagations, Random Oper. Stoch. Equ. (March 2015), vol 23, Issue 1 DOI:10.1515/rose-2014-0029.

\end{itemize}

\end{document}